# Reduction of Ordered Spin Moments in Antiferromagnets of $S = 5/2$ Ions ($Fe^{3+}$, $Mn^{2+}$) Driven by Local Magnetic Excitation

Myung-Hwan Whangbo[1,*], Reinhard K. Kremer[2] and Hyun-Joo Koo[3,*]

[1] Department of Chemistry, North Carolina State University, Raleigh, NC 27695-8204, USA

[2] Max Planck Institute for Solid State Research, Heisenbergstrasse 1, D-70569 Stuttgart, Germany

[3] Department of Chemistry, Research Institute for Basic Sciences, Kyung Hee University, Seoul 02447, Republic of Korea

**Abstract**

For antiferromagnets composed of $S = 5/2$ ions such as $Fe^{3+}$ and $Mn^{2+}$, the moments of these ions in the ordered antiferromagnetic state can be significantly smaller than 5 $\mu_B$ (i.e., 1.56 – 4.48 $\mu_B$) if these ions form quantum fluctuating entities (QFEs), e.g., quasi-one-dimensional uniform antiferromagnetic chains or quasi-zero-spin antiferromagnetic spin dimers. It is reasonable to suppose that the stronger the quantum fluctuation in such an antiferromagnet, the greater the reduction in its ordered moment would become, but this supposition has not yet been confirmed because quantifying the strength of quantum fluctuation is not a straightforward matter. Here we show that the local magnetic excitations involving the QFEs can be used to quantify the strength of quantum fluctuation by analyzing six antiferromagnets showing significant reduction in their ordered spin moments.

mike_whangbo@ncsu.edu

hjkoo@khu.ac.kr

For antiferromagnets composed of $S = 5/2$ ions undergoing a long-range antiferromagnetic (AFM) ordering, the spin moments $\mu_S$ of these ions are expected to be close to 5 $\mu_B$ in the ordered AFM state at very low temperatures. For such ions, spin-orbit coupling is negligible,[1] making their spin exchange coupling isotropic (i.e., the Heisenberg-type). For antiferromagnets such as $LiFeGe_2O_6$, $NaFeGe_2O_6$, $SrMn_2V_2O_8$, $FeVMoO_7$, $LiFeV_2O_7$ and $Cu_2Fe_2Ge_4O_{13}$, however, the ordered spin moments $\mu_S$ are considerably smaller than 5 $\mu_B$ (i.e., 1.56 – 4.48 $\mu_B$)[2,3] in the absence of strong spin frustration. This has been understood as a consequence of quantum fluctuation arising from the presence of such QFEs as quasi-one-dimensional (Q1D) uniform AFM chains[2,4] or quasi-zero-spin (Q0S) spin dimers[2,3] in the antiferromagnets, in which each QFE interacts weakly with its surrounding QFEs. A weaker reduction in the ordered spin moments of such an antiferromagnet is expected if the quantum fluctuation of its QFEs becomes weaker in strength. To confirm this supposition, we quantify the strength of quantum fluctuation in a given antiferromagnet by analyzing the energy required for its local magnetic excitation,[3,5] namely, the energy needed to isolate one QFE from its host antiferromagnet without affecting all the remainder of the host. In probing this question, we will employ the concept of magnetic bonds;[5,6] a strong (weak) magnetic bond refers to a strong (weak) AFM spin exchange path in which two magnetic ions are antiferromagnetically coupled, and a magnetic "antibond" to an AFM spin exchange path in which the two magnetic ions are ferromagnetically coupled. For a ferromagnetic (FM) spin exchange path, however, its magnetic bond (antibond) is defined to have the two magnetic ions ferromagnetically (antiferromagnetically) coupled.

The collinearly ordered AFM state of an antiferromagnet can be represented by the "↑↓↑↓⋯" or "↓↑↓↑⋯" spin arrangement because the two are equal in energy. Similarly, the cycloid spin arrangement[7] of an antiferromagnet can be represented by a right- or left-handed cycloid because the two are equal in energy. Studies on single crystal antiferromagnets in their ordered AFM (cycloid) state show[8] that domains of the "↑↓↑↓⋯" and "↓↑↓↑⋯" (right- and left-handed cycloid) spin arrangements coexist. In an antiferromagnet of Q0S spin dimers, these QFEs are surrounded only with weaker magnetic bonds. In a magnetic domain (say, domain A) where the ↑↓ spin arrangement (Figure 1a) for the strongest magnetic bond is pinned, its alternative ↓↑ spin arrangement would be slightly less stable. The preferential pinning of one spin arrangement is most likely due to a trace of crystal defects. The opposite is the case in a domain where the ↓↑ arrangement (say, domain B) (Figure 1b) for the strongest magnetic bond is pinned. In either case, spin flipping of the strongest magnetic bond (i.e., an example of local magnetic excitation) breaks several weaker magnetic bonds connected to it, hence requiring a certain amount of activation energy, $E_a$. For an antiferromagnet of Q1D uniform chains (Figure 1c), the interchain magnetic bonds are usually much weaker than the intrachain nearest-neighbor (NN) magnetic bond. This makes it unlikely to spin flip one NN magnetic bond within each uniform AFM chain because it requires breaking two NN magnetic bonds. To isolate one Q1D uniform AFM chain from its host antiferromagnet, all its weak interchain magnetic bonds should be broken. This requires a certain activation energy $E_a$ per NN magnetic bond of the chain. In an isolated 1D uniform AFM Heisenberg chain, long-range AFM ordering is absent. Instead, short-range order (SRO) fluctuations produce dynamic clusters of AFM spin arrangements (domains A and B) throughout the chain (Figure 1d). Every dynamic AFM cluster is separated from each other by a broken NN magnetic bond at each end by the very nature of SRO. Then, isolating a Q1D uniform AFM chain from the host antiferromagnet amounts to introducing local magnetic excitations in the form of SRO clusters, the activation energies $E_a$ for which do not involve breaking the NN magnetic bonds.

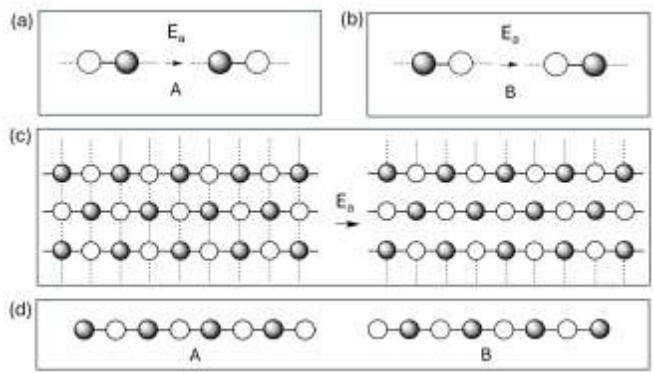

Figure 1. (a, b) Local excitations in antiferromagnets made up of Q0S AFM spin dimers. The strongest magnetic bond has the ↑↓ spin arrangement in (a), but the ↓↑ spin arrangement in (b). (c) Isolating one 1D uniform AFM chain from an antiferromagnet made up of such chains by breaking all its interchain magnetic bonds. (d) Dynamic clusters with the "↑↓↑↓…" and "↓↑↓↑…"spin arrangements, which occur in an isolated 1D uniform AFM chain due to SRO fluctuations. Here the empty and filled spheres represent the up-spins and down-spins, respectively.

In this work, we examine how the activation energies $E_a$ are related to the reduction in the ordered spin moments of the antiferromagnets $LiFeGe_2O_6$,[9] $NaFeGe_2O_6$,[10] $SrMn_2V_2O_8$,[4] $FeVMoO_7$,[2] and $LiFeV_2O_7$,[3] each of which has only one kind of magnetic $S = 5/2$ ions. We evaluate the spin exchanges of these antiferromagnets using the energy-mapping analysis[11] based on DFT+U calculations, find their spin lattices, identify the local magnetic excitations, and finally examine how these local magnetic excitations are related to the reduced ordered moments and hence to the extent of quantum fluctuation. We evaluate the spin exchanges of $LiFeGe_2O_6$, $NaFeGe_2O_6$, $SrMn_2V_2O_8$, and $FeVMoO_7$ by performing the energy mapping analyses (see Section S1 of the supporting information, SI) based on DFT+U calculations using the spin Hamiltonian defined as

$$H_{spin} = \sum_{i>j} J_{ij} \vec{S}_i \cdot \vec{S}_j, \tag{1}$$

so that AFM and FM spin exchanges are represented by $J_{ij} > 0$ and $J_{ij} < 0$, respectively. The spin exchanges of $LiFeV_2O_7$ have already been reported.[4] Once the spin exchanges of a given magnet are available, we illustrate how to identify the relevant local excitation and calculate the associated activation energy $E_a$ by considering one representative example, $FeVMoO_7$. Our analysis for the remaining antiferromagnets are presented in Section S2 in the SI. Details of DFT+U calculations[12-15] and our energy-mapping analyses are summarized in Section S3 of the SI.

$FeVMoO_7$ consists of $FeO_6$ octahedra containing $Fe^{3+}$ ions. Every two $FeO_6$ octahedra share their edges to form a $Fe_2O_{10}$ dimer (Figure 2a) leading to the Fe-O-Fe exchange path $J_1$. These $Fe_2O_{10}$ dimers are interconnected through $MoO_4$ and/or $VO_4$ tetrahedra by corner sharing to form Fe-O…$A^{n+}$…O-Fe exchange paths $J_2 - J_5$, where $A^{n+} = Mo^{6+}$ or $V^{5+}$.[11,16] The arrangement of $J_1 - J_5$ bonds is presented in Figure 2b by showing only the $Fe^{3+}$ ions, where the $J_1$ and $J_4$ bonds are explicitly connected. Since the oxygen positions of the reported crystal structure are not accurate enough, we optimized the crystal structure and determined the values of $J_1 - J_5$ using the experimental and the optimized crystal structures. The Fe…Fe distances associated with $J_1 - J_5$ are presented in Figure 2c, and so are the $J_1 - J_5$ values obtained from the optimized structure. The most stable spin arrangement based on these spin exchanges (Figure 2d) shows that the alternating AFM chains defined by $J_1$ and $J_4$ are spin frustrated by the interchain spin exchanges $J_2$, $J_3$ and $J_5$.

When one strongest bond $J_1$ spin-flips, two $J_4$ and four $J_5$ bonds are broken while two $J_2$ and two $J_3$ "antibonds" are removed, so that $E_a = (2J_4 + 4J_5 - 2J_2 - 2J_3)S^2 = 6.6\,S^2 = 41.3$ K. For comparing all five different antiferromagnets on an equal footing, we define the activation index $\delta_a$ as

$$\delta_a = E_a/J_m \qquad (2)$$

where $J_m$ is the strongest spin exchange (i.e., $J_1$ in the present case) for each antiferromagnet. The activation indices calculated for the five antiferromagnets are summarized in Table 1.

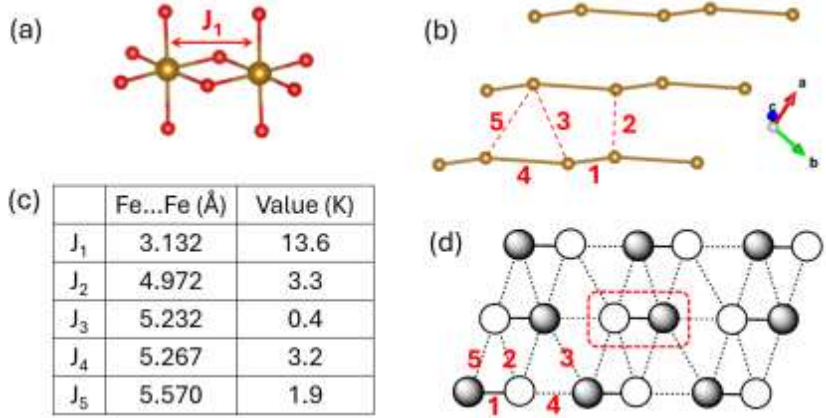

Figure 2. (a) Edge-sharing dimer $Fe_2O_{10}$ of $FeVMoO_7$. (b) Spin lattice of $FeVMoO_7$ defined by five spin exchange paths $J_1 - J_5$. (c) Fe…Fe distances associated with the calculated spin exchanges for $J_1 - J_5$. (d) The most stable spin arrangement for the spin lattice of $FeVMoO_7$. In (b) and (d), the numbers 1 – 5 refer to the exchange paths $J_1 - J_5$, respectively.

**Table 1.** Table 1. Ordered spin moments $\mu_s$ (in $\mu_B$) measured at $T_{mes}$ (in K), effective energy gaps $\Delta_{eff}$ (in K), strongest magnetic bond $J_m$ (in K), activation energies $E_a$ (in K), activation indices $\delta_a$, and Néel temperature (in K) of five antiferromagnets.

|  | $\mu_s$ | $T_{mes}$ | $\Delta_{eff}$ | $J_m$ | $E_a$ | $\delta_a$ | $T_N$ |
|---|---|---|---|---|---|---|---|
| LiFeGe$_2$O$_6$ | 4.48(5) | 1.5 | 11.3 | 12.5 | 43.6 | 3.49 | 20.2 |
| NaFeGe$_2$O$_6$ | 4.09(4) | 2.5 | 4.26 | 19.0 | 2.5 | 0.13 | 11.2 |
| SrMn$_2$V$_2$O$_8$ | 3.99(1) | 1.5 | 2.40 | 44.3 | 20.5 | 0.46 | 42.2 |
| FeVMoO$_7$ | 4.00(7) | 4 | 6.40 | 13.6 | 41.3 | 3.03 | 10.8 |
| LiFeV$_2$O$_7$ | 3.41(1)[a] | 1.5 | 1.72 | 93.1 | 68.1 | 0.73 | ~5 |

[a] For the Fe2 atom of LiFeV$_2$O$_7$. For the ordered spin moments of Fe1 and Fe3, see the text.

As already pointed out, in each magnetic domain A, the ↑↓ spin arrangement is slightly more stable than the ↓↑ spin arrangement (Figure 1a) due to the preferential pinning of the ↑↓ spin arrangement. Thus, the ↑↓ and ↓↑ spin arrangements of domain A can be regarded as the ground state $\psi_G$ and the excited state $\psi_E$, with small energy gap $\Delta$ (in $k_B$K units) between the two (Figure 3a, Left). Although $\Delta$ is small, the Boltzmann factor $\lambda$ at $T_{mes}$,

$$\lambda = \exp\left(-\frac{\Delta}{T_{mes}}\right), \tag{3}$$

cannot be neglected because the measuring temperatures $T_{mes}$ (e.g., for the neutron diffraction experiment, they are very low, Table 1). In this two-state approximation, the excited state $\psi_E$ becomes populated by the amount of $\lambda$, so that the resulting spin moment is given by $\mu_G + \lambda\mu_E = (1 - \lambda)\mu_G$, where $\mu_G$ and $\mu_E$ represent the moments associated with the ↑↓ and ↓↑ spin arrangements, respectively (Figure 3a, Right). Similarly, the same moment reduction occurs in each magnetic domain B (Figure 3b) and so does in each dynamic cluster generated by the SRO fluctuations in an isolated 1D uniform AFM chain (Figure 3c,d).

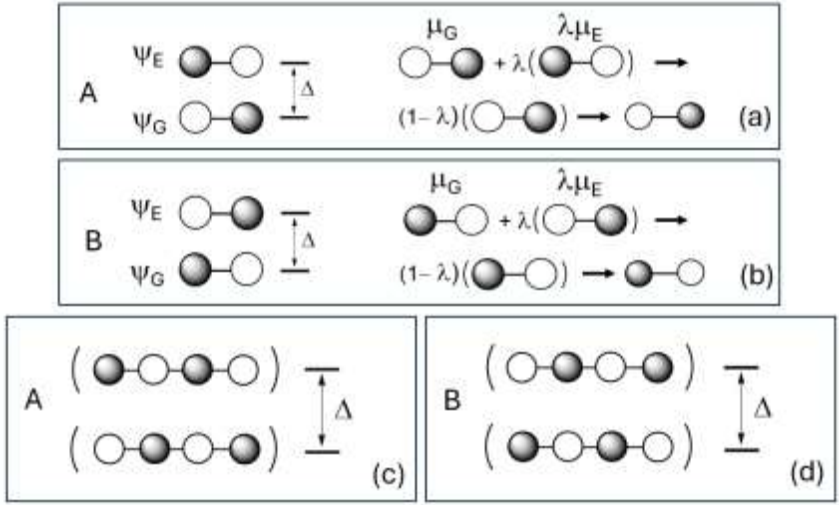

Figure 3. (a, b) Spin arrangements describing the ground and excited states associated with the strongest magnetic bond in an antiferromagnet composed of Q0S AFM spin dimers. (c, d) Spin arrangements describing the ground and excited states associated with the dynamic clusters (indicated by brackets), which are generated by SRO fluctuations in a 1D uniform AFM chain.

The extent of moment reduction per 1 $\mu_B$ for a magnetic ion of spin $S$ is written as $(2S - \mu_s)/2S$. To a first approximation, the latter can be equated to the Boltzmann factor $\lambda$ at $T_{mes}$ (Eq. 3). Then, the effective energy gap $\Delta_{eff}$ can be estimated as follows:

$$\Delta_{eff} = -T_{mes} \ln\left(\frac{2S-\mu_s}{2S}\right) \qquad (4)$$

The observed values of $T_{mes}$ and $\mu_s$ as well as the calculated values of $\Delta_{eff}$ by using Eq. 4 are presented in Table 1, and the $\delta_a$ vs. $\Delta_{eff}$ plot in Figure 4. For the antiferromagnets composed of $Fe^{3+}$ ions, $\Delta_{eff}$ increases with increasing $\delta_a$ for both the Q1D uniform AFM chain and the Q0S AFM spin-dimer systems, and the $\delta_a$ vs. $\Delta_{eff}$ plots for the two systems have nearly the same slope. The $\delta_a$ vs. $\Delta_{eff}$ plot for the Q0S AFM spin-dimer systems is described by $\delta_a \approx 0.5 \Delta_{eff}$, and that for the Q1D uniform AFM chain systems by $\delta_a \approx 0.5 \Delta_{eff} - 1.8$. The point for $SrMn_2V_2O_8$ deviates somewhat from the correlation line of the $\delta_a$ vs. $\Delta_{eff}$ plot found for the Q1D uniform AFM chain systems probably due to the difference in the magnetic ions (i.e., $Mn^{2+}$ vs. $Fe^{3+}$). Certainly, more examples are needed to confirm the trend satisfactorily, but it is strongly suggested that the smaller the activation index (i.e., the normalized activation energy for the local magnetic excitation), the stronger the quantum fluctuation in the QFE hence inducing a greater reduction of the ordered spin moment in each QFE.

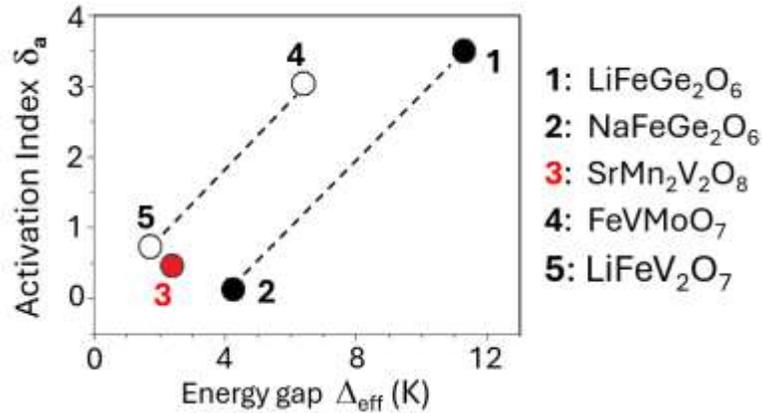

Figure 4. Plot of the activation energy $\delta_a$ vs. the effective energy gap $\Delta_{eff}$ determined for five antiferromagnets. The filled circles denote three antiferromagnets consisting of Q1D uniform AFM chains, while the empty circles indicate those comprised of Q0S AFM spin dimers. The dashed lines are only the guide for the eye.

For a given antiferromagnet, the activation index $\delta_a$ of the QFE is determined by its spin exchanges, whereas the effective energy gap $\Delta_{eff}$ of the QFE is determined by its ordered spin moments determined from neutron diffraction measurements at $T_{mes}$, assuming that the ground and locally excited states have an energy gap of $\Delta_{eff}$. Simply speaking, $\delta_a$ is needed to isolate a QFE from its host antiferromagnet, while $\Delta_{eff}$ determines how strongly the locally excited state is populated by the Boltzmann factor hence reducing the ordered moment of each QFE.

Our discussion of local magnetic excitations implicitly assumed that they generate identical QFEs. LiFeV$_2$O$_7$ consists of distorted diamond chains of Fe atoms in which Fe2 monomer units alternate with perpendicular Fe1-Fe3 dimer units (Section S2).[3] At 1.5 K, the ordered spin moment is 1.56 $\mu_B$ for the Fe1 and Fe3 atoms, much smaller than 3.41 $\mu_B$ found for Fe2. As already pointed out,[3] the local excitation (i.e., spin flip) of each Fe1-Fe3 dimer unit requires the same activation energy $E_a$ as does that of each Fe2 monomer unit (Section S2), leading to the ordered moment of 3.41 $\mu_B$ to all three Fe atoms. Each Fe1-Fe3 dimer is a Q0S AFM spin dimer whereas each Fe2 monomer is an $S = 5/2$ entity, so quantum fluctuations that occur in each Fe1-Fe3 dimer does not in each Fe2 monomer. This explains why the ordered spin moment of Fe1 and Fe3 is further reduced to 1.56 $\mu_B$.

In conclusion, the extent of quantum fluctuation in antiferromagnets made up of quasi-1D uniform antiferromagnetic chains and quasi-zero-spin spin dimers can be quantified in terms of the normalized activation energies associated with their local magnetic excitations.

**Supporting Information**

Supplementary Section S1 – S3 with Fig. S1 – S9 and Table S1 – S16 (calculations of $E_a$, $\delta_a$ and $\Delta_{eff}$ as well as results of energy mapping analysis). All are in pdf.

**Acknowledgements**

The research at KHU was supported by Basic Science Research Program through the National Research Foundation of Korea (NRF) funded by the Ministry of Education (RS-2020-NR049601).

TOC figure

For Table of Contents Only

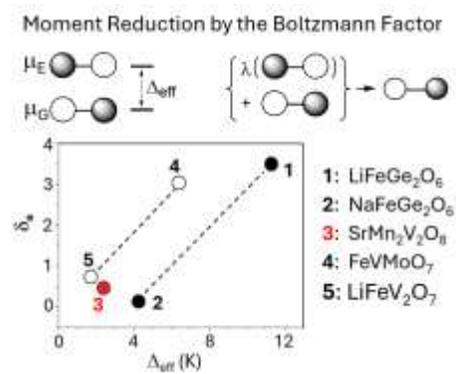

Supporting information

for

**Reduction of Ordered Spin Moments in Antiferromagnets of $Fe^{3+}$ ($S$ = 5/2) Ions Driven by Local Magnetic Excitation**

Myung-Hwan Whangbo[1,*], Reinhard K. Kremer[2] and Hyun-Joo Koo[3,*]

[1] Department of Chemistry, North Carolina State University, Raleigh, NC 27695-8204, USA

[2] Max Planck Institute for Solid State Research, Heisenbergstrasse 1, D-70569 Stuttgart, Germany

[3] Department of Chemistry, Research Institute for Basic Sciences, Kyung Hee University, Seoul 02447, Republic of Korea

mike_whangbo@ncsu.edu

hjkoo@khu.ac.kr

## S1. Energy-mapping analysis for spin exchanges

For an antiferromagnet to be described by N different spin exchanges, $J_1 - J_N$, we construct N+1 ordered spin states to obtain their energies $E_{SE}(i)$ (i = 1, 2, …, N+1) written in terms of $J_1 - J_N$.

$$E_{SE}(i) = \sum_{i=1}^{N} n_1(i) J_1 + n_2(i) J_2 + \cdots + n_N(i) J_N. \tag{S1}$$

From this one obtains their N relative energies relative energies $\Delta E_{SE}(i)$ (i = 1 − N), which are written in terms of $J_1 - J_N$. To obtain the numerical values of $J_1 - J_N$, we calculate the energies of N+1 ordered spin states to obtain their energies $E_{DFT}(i)$ (i = 1, 2, …, N+1) and hence N relative energies $\Delta E_{DFT}(i)$ (i = 1 − N). Then, by equating $\Delta E_{SE}(i)$ (i = 1 − N) to $\Delta E_{DFT}(i)$ (i = 1 − N), the numerical values of $J_1 - J_N$ are obtained. We carried out DFT calculations using the frozen core projector augmented plane wave (PAW) [12] encoded in the Vienna ab initio Simulation Packages (VASP) [13] and the PBE functional [14] for the exchange-correlation functional. To take into consideration the electron correlation associated with the 3*d* states of Fe and Mn, we perform DFT+*U* calculations[15] using an effective on-site repulsion $U_{eff} = U - J = 3$ and 4 eV. In the main text and in Section S2, we present our analyses based on the J values obtained with $U_{eff} = 4$ eV, because the J values obtained with $U_{eff} = 3$ eV have the same trends as do those obtained from those with $U_{eff} = 4$ eV (Section S3).

## S2. Analysis of local magnetic excitation

### A. AFeGe$_2$O$_6$ (A = Na, Li)

AFeGe$_2$O$_6$ (A = Na, Li) consists of zigzag chains running along the c-direction, which are made up of *cis*-edge-sharing FeO$_6$ octahedra (**Fig S1a**). Such zigzag chains repeat along the a- and b-directions (**Fig. S1b**). We describe each zigzag chain by using the nearest-neighbour (NN) spin exchange $J_1$, which is the Fe-O-Fe type spin exchange. Between adjacent zigzag chains of $J_1$, there occur four Fe-O…O-Fe type spin exchanges $J_2$, $J_3$, $J_2'$, and $J_3'$ (**Fig. S1b**). The three spin exchanges $J_1$, $J_2$ and $J_3$ form the "horizontal" corrugated layers (**Fig. S1b,c**) parallel to the (a-b)c-plane. The three spin exchanges $J_1$, $J_2'$ and $J_3'$ form the "perpendicular" corrugated layers (**Fig. S1b,d**) parallel to the (a+b)c-plane.

Every corrugated layer, be it "horizontal" or "perpendicular", can be regarded as made up of two kinds of ladders defined by $J_1$ and $J_3$. The ladders A have $J_2$'s, but the ladders B do not. The "horizontal" and "perpendicular" corrugated layers are condensed together such that each zigzag chain of every "horizontal" layer is the same as that of every "perpendicular" layer, and such that the ladders A(B) of each "horizontal" layer are condensed with the ladders B(A) of the "perpendicular" layer. Consequently, to isolate one zigzag chain from the resulting three-dimensional (3D) spin lattice, it necessary to break four $J_3$ and two $J_2$ per $J_1$ from the horizontal layer as well as four $J_3'$ and two $J_2'$ per $J_1$ from the perpendicular layer. For NaFeGe$_2$O$_6$, the "horizontal" layers are identical in structure to the "vertical" layers so that $J_2 = J_2'$, and $J_3 = J_3'$. For LiFeGe$_2$O$_6$, however, this is not the case.

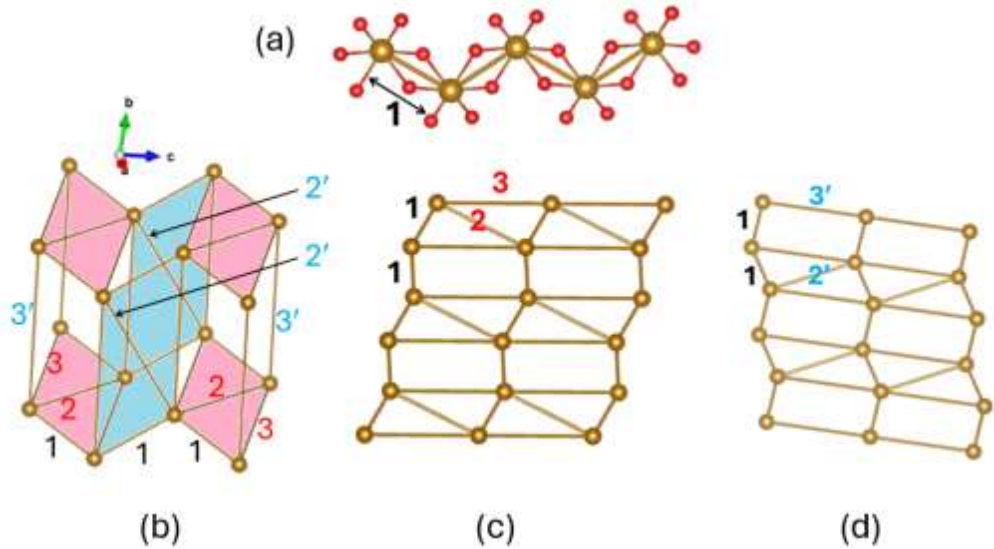

**Fig. S1**. (a) Zigzag chain of *cis*-edge-sharing $FeO_6$ octahedra. (b) Fe-O…O-Fe type spin exchanges $J_2$, $J_3$, $J_2'$, and $J_3'$ that occur between adjacent zigzag chains of $J_1$. (c) Horizontal corrugated layer made up of $J_1$, $J_2$ and $J_3$. (d) Perpendicular corrugated layer made up of $J_1$, $J_2'$ and $J_3'$.

The spin exchanges $J_1 - J_3$ calculated for $NaFeGe_2O_6$ are summarized in **Fig. S2a**, and the most stable spin arrangement predicted for each layer (either "horizontal" or "vertical") is presented in **Fig. S2b**. Per $J_1$ bond of the uniform chain, four $J_3$ bonds should be broken to isolate a chain. This leads to the breaking of two $J_2$ antibonds. Thus, $E_a = (4J_3 - 2J_2) S^2 = 0.4\ S^2 = 2.5$ K using the spin exchanges obtained for the optimized structure.

The spin exchanges $J_1$, $J_2$, $J_3$, $J_2'$, and $J_3'$ calculated for $LiFeGe_2O_6$ are summarized in **Fig. S3a**. The most stable spin arrangement predicted for the "horizontal" is presented in **Fig. S3b**, and that for the "vertical" layer in **Fig. S3c**. In the horizontal layer, per NN bond $J_1$, four $J_3$ bonds and two $J_2$ antibonds are broken. Thus, $E_{a,\ horizontal} = (4J_3 - 2J_2) S^2 = 7.22\ S^2 = 45.1$ K. In the vertical layer, per NN bond $J_1$, isolating one chain involves the breaking of two $J_2$ bonds and four $J_3$ antibonds. Thus, $E_{a, \text{vertical}} = (-4J_3' + 2J_2')S^2 = 6.74 S^2 = 42.1$ K. On average, $E_a = (E_{a, \text{horizontal}} + E_{a, \text{vertical}})/2 = 43.6$ K.

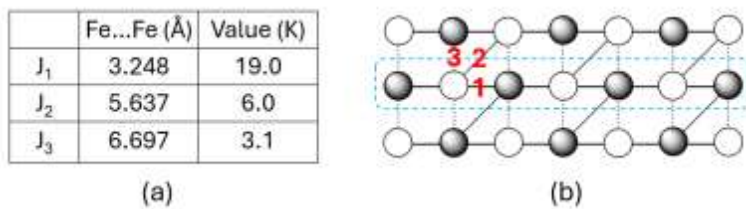

|  | Fe...Fe (Å) | Value (K) |
|---|---|---|
| $J_1$ | 3.248 | 19.0 |
| $J_2$ | 5.637 | 6.0 |
| $J_3$ | 6.697 | 3.1 |

(a)  (b)

Fig. S2. (a) The Fe…Fe distances associated with and the values of $J_1 - J_3$ determined for the optimized structure of NaFeGe$_2$O$_6$. (b) The most stable spin arrangement for the horizontal and vertical layers of NaFeGe$_2$O$_6$.

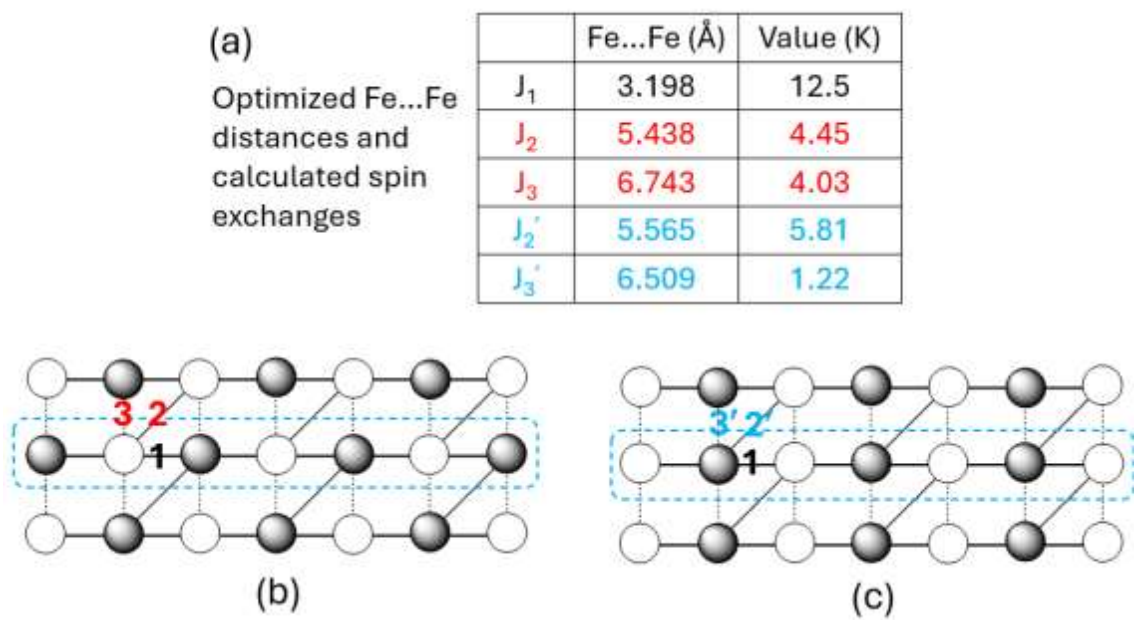

(a) Optimized Fe...Fe distances and calculated spin exchanges

|  | Fe...Fe (Å) | Value (K) |
|---|---|---|
| $J_1$ | 3.198 | 12.5 |
| $J_2$ | 5.438 | 4.45 |
| $J_3$ | 6.743 | 4.03 |
| $J_2'$ | 5.565 | 5.81 |
| $J_3'$ | 6.509 | 1.22 |

(b)  (c)

Fig. S3. (a) The Fe…Fe distances associated with and the values of $J_1$, $J_2$, $J_3$, $J_2'$, and $J_3'$ determined for the optimized structure of LiFeGe$_2$O$_6$. (b) The most stable spin arrangement for the horizontal layers of NaFeGe$_2$O$_6$. (c) The most stable spin arrangement for the vertical layers of NaFeGe$_2$O$_6$.

**B. SrMn$_2$V$_2$O$_8$**

SrMn$_2$V$_2$O$_8$ consists of spiral chains running along the c-direction, which are made up of *cis*-edge-sharing MnO$_6$ octahedra (**Fig S4a**). Each spiral chain is condensed with VO$_4$ tetrahedra. Such spiral chains repeat along the a- and b-directions (**Fig. S4b**). We describe each zigzag chain by using the NN spin exchange $J_1$, which is the Fe-O-Fe type spin exchange, as well as the next-nearest-neighbour (NNN) spin exchange $J_2$, which is the Fe-O…O-Fe type spin exchange (**Fig. S4c**). Between adjacent zigzag chains, there occur the Fe-O…O-Fe type spin exchange $J_3$. The spin exchanges $J_1$, $J_2$ and $J_3$ calculated for SrMn$_2$V$_2$O$_8$ are summarized in **Fig. S4d**. It is clear from **Fig.S4b** and **Fig. S4c** that, to isolate one spiral chain, two inter-chain bonds $J_3$ should be broken per NN bond $J_1$ of the spiral chain. Therefore, $E_a = 2J_3 = 3.28\ S^2 = 20.5$ K.

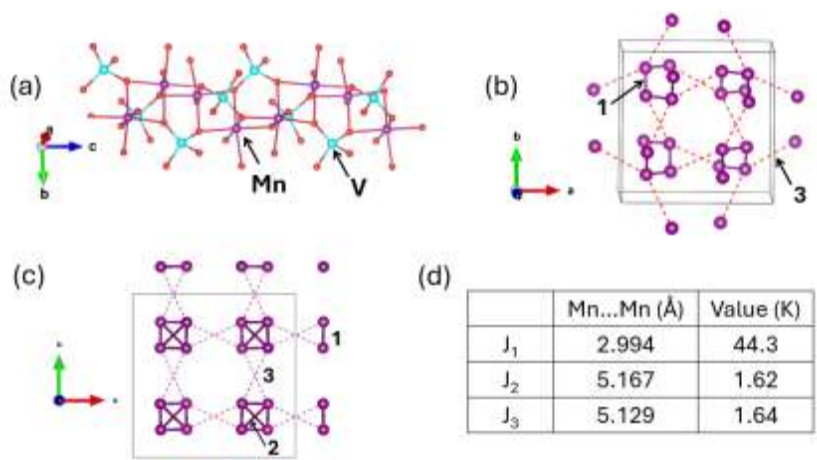

**Fig. S4**. (a) One spiral chain of $SrMn_2V_2O_8$ condensed with $VO_4$ tetrahedra. (b) A projection view of spiral chains along the direction slightly away from the c-axis. (c) A projection view of spiral chains along the c-axis. (d) The Mn…Mn distances associated with and the values of $J_1 - J_3$ determined for the optimized structure of $SrMn_2V_2O_8$.

## C. LiFeV$_2$O$_7$

Since $LiFeV_2O_7$ was studied in ref. 3, we briefly summarize the essential results. **Fig. S5a** how one diamond chain is surrounded by four diamond chains. Each monomer (Fe2) makes two inter-chain bonds $J_6$ and $J_7$, and so does each Fe1-Fe3 dimer. The most stable spin arrangement of a diamond chain is depicted in **Fig. S5b**, where the shaded and unshaded spheres represent the down-spin and up-spins, respectively, which is based on the spin exchanges of $J_1 - J_7$ listed in **Fig. S5c**. The ordered spin moments of Fe1, Fe2 and Fe3 determined by neutron diffraction at 1.5 K are given in **Fig. S5d**. Spin flip of a monomer unit Fe2 breaks two intra-chain bonds (i.e., $J_3$ and $J_4$), two intra-chain antibonds (i.e., $J_1$ and $J_2$) and two inter-chain bonds (i.e., $J_6$ and $J_7$). Thus, the activation energy for the spin flip of each Fe2 is given by $E_a = (J_3 + J_4 - J_1 - J_2 + J_6 + J_7)S^2 = 10.9 S^2 = 68.1$ K. The same activation energy is obtained for the spin flip of each Fe1-Fe3 dimer.

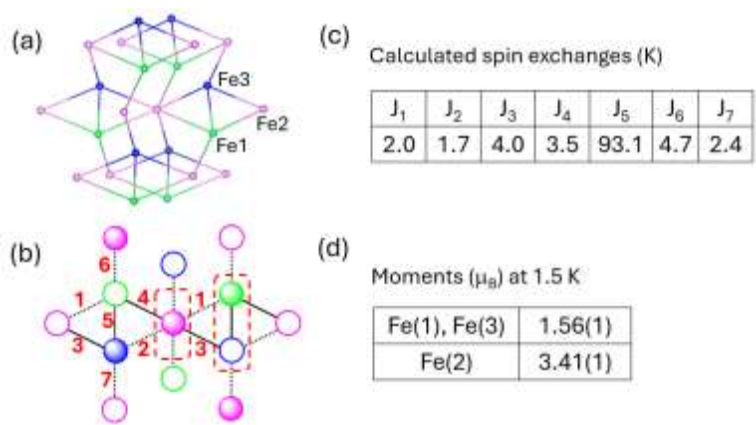

**Fig. S5**. (a) Four diamond chains (represented by one diamond per chain for simplicity) surrounding one diamond chain in LiFeV$_2$O$_7$. (b) The most stable spin arrangement of diamond chains with shaded and unshaded spheres representing the down-spin and up-spins, respectively. (c) The values of the spin exchanges $J_1 - J_7$. (d) The ordered spin moments of Fe1, Fe2 and Fe3 determined by neutron diffraction at 1.5 K.

## S3. Computational details

All our DFT+U calculations employed the plane wave cutoff energy of 450 eV, the threshold of $10^{-6}$ eV for self-consistent-field energy convergence. For our crystal structure optimizations, the space group and the cell parameters of each antiferromagnet were fixed as found experimentally, but the atom positions were relaxed with the criterion of 0.03eV/Å. The k-point sets used are (8×8×6), (4×4×6), (4×4×8) and (4×4×6) for $FeVMoO_7$, $NaFeGe_2O_6$, $LiFeGe_2O_6$ and $SrMn_2V_2O_8$, respectively.

### 1. FeVMoO$_7$

a) Table S1. The optimized atom positions of FeVMoO$_7$

| Atom | x | y | z |
|---|---|---|---|
| Fe1 | 0.086782 | 0.346289 | 0.594741 |
| Fe2 | 0.413218 | 0.153711 | 0.405259 |
| V1 | 0.344281 | 0.378959 | 0.334791 |
| V2 | 0.155719 | 0.121041 | 0.665209 |
| Mo1 | 0.146847 | 0.104481 | 0.108552 |
| Mo2 | 0.353153 | 0.395519 | 0.891448 |
| O1 | 0.445497 | 0.312321 | 0.423887 |
| O2 | 0.054503 | 0.187679 | 0.576113 |
| O3 | 0.104561 | 0.995812 | 0.609302 |
| O4 | 0.395439 | 0.504188 | 0.390698 |
| O5 | 0.387287 | 0.026963 | 0.901099 |
| O6 | 0.112713 | 0.473037 | 0.098901 |
| O7 | 0.333000 | 0.351961 | 0.106641 |
| O8 | 0.167000 | 0.148039 | 0.893359 |
| O9 | 0.204615 | 0.344611 | 0.412206 |
| O10 | 0.295385 | 0.155389 | 0.587794 |
| O11 | 0.216953 | 0.351130 | 0.766023 |
| O12 | 0.283047 | 0.148870 | 0.233977 |
| O13 | 0.470316 | 0.344672 | 0.785012 |
| O14 | 0.029684 | 0.155328 | 0.214988 |

b) The ordered spin states of FeVMoO$_7$.

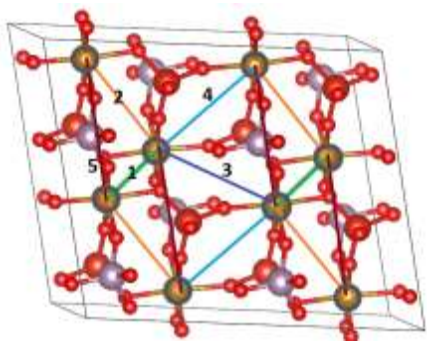
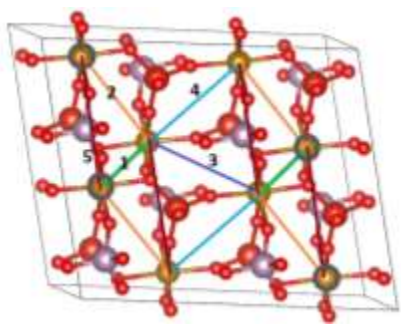

(a) FM                                              (b) AF1

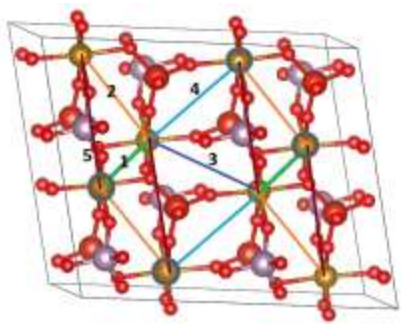
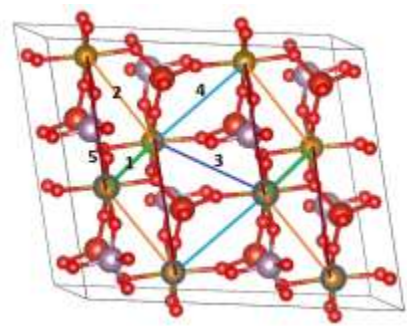

(c) AF2                                              (d) AF3

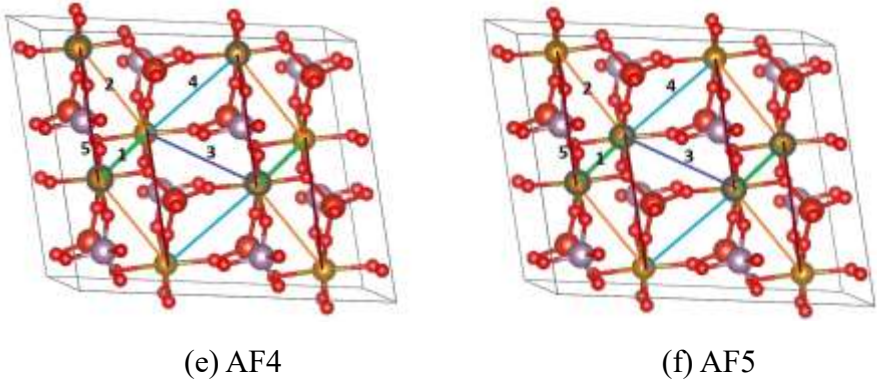

(e) AF4            (f) AF5

**Fig. S6**. The ordered spin states used for FeVMoO$_7$: (a) FM, (b) AF1, (c) AF2, (d) AF3, (e) AF4, and (f) AF5 states, where up-spin Fe$^{3+}$ sites are indicated by encircling them with black circles.

Table S2. The spin exchange energies for the ordered spin states of FeVMoO$_7$.

$E_{SE}(i) = \sum_{i=1}^{N} n_1(i) J_1 + n_2(i) J_2 + \cdots + n_N(i) J_N$.

| i | $n_1$ | $n_2$ | $n_3$ | $n_4$ | $n_5$ |
|---|---|---|---|---|---|
| FM | +4 | +4 | +4 | +4 | +8 |
| AF1 | -4 | -4 | +4 | +4 | +8 |
| AF2 | -4 | +4 | +4 | -4 | -8 |
| AF3 | -4 | +4 | -4 | +4 | -8 |
| AF4 | -4 | -4 | -4 | -4 | +8 |
| AF5 | +4 | -4 | +4 | -4 | -8 |

c) Table S3. The relative energies of the ordered spin states in terms of DFT+U calculations for the optimized structure of FeVMoO$_7$.

| i | U = 3 eV | U = 4 eV |
|---|---|---|
| FM | 14.45 | 10.91 |
| AF1 | 2.88 | 2.00 |
| AF2 | 0.00 | 0.00 |
| AF3 | 2.18 | 1.55 |
| AF4 | 0.07 | 0.04 |
| AF5 | 6.79 | 5.36 |

d) Table S4. The values of the spin exchanges determined for $FeVMoO_7$ by the energy mapping analysis

| | U = 3 eV | U = 4 eV |
|---|---|---|
| $J_1$ | 17.04 | 13.24 |
| $J_2$ | 4.44 | 3.29 |
| $J_3$ | 0.59 | 0.39 |
| $J_4$ | 4.63 | 3.25 |
| $J_5$ | 2.58 | 1.88 |

## 2. $NaFeGe_2O_6$

a) Table S5. The optimized atom positions of $NaFeGe_2O_6$.

| | x | y | z |
|---|---|---|---|
| Fe1 | 0.000000 | 0.904908 | 0.250000 |
| Fe2 | 0.000000 | 0.095092 | 0.750000 |
| Fe3 | 0.500000 | 0.404908 | 0.250000 |
| Fe4 | 0.500000 | 0.595092 | 0.750000 |
| Na1 | 0.000000 | 0.301458 | 0.250000 |
| Na2 | 0.000000 | 0.698542 | 0.750000 |
| Na3 | 0.500000 | 0.801458 | 0.250000 |
| Na4 | 0.500000 | 0.198542 | 0.750000 |
| Ge1 | 0.288430 | 0.094852 | 0.229081 |
| Ge2 | 0.711570 | 0.905148 | 0.770919 |
| Ge3 | 0.711570 | 0.094852 | 0.270919 |
| Ge4 | 0.288430 | 0.905148 | 0.729081 |
| Ge5 | 0.788431 | 0.594853 | 0.229082 |
| Ge6 | 0.211569 | 0.405147 | 0.770918 |
| Ge7 | 0.211569 | 0.594853 | 0.270918 |
| Ge8 | 0.788431 | 0.405147 | 0.729082 |
| O1 | 0.103955 | 0.083105 | 0.131585 |
| O2 | 0.896045 | 0.916895 | 0.868415 |
| O3 | 0.896045 | 0.083105 | 0.368415 |
| O4 | 0.103955 | 0.916895 | 0.631585 |
| O5 | 0.603953 | 0.583105 | 0.131584 |
| O6 | 0.396046 | 0.416895 | 0.868416 |
| O7 | 0.396046 | 0.583105 | 0.368416 |
| O8 | 0.603954 | 0.416895 | 0.631584 |
| O9 | 0.355927 | 0.274714 | 0.303477 |
| O10 | 0.644073 | 0.725286 | 0.696524 |
| O11 | 0.644073 | 0.274714 | 0.196524 |
| O12 | 0.355927 | 0.725286 | 0.803476 |
| O13 | 0.855927 | 0.774713 | 0.303476 |
| O14 | 0.144073 | 0.225287 | 0.696524 |
| O15 | 0.144073 | 0.774713 | 0.196524 |
| O16 | 0.855927 | 0.225287 | 0.803476 |
| O17 | 0.365433 | 0.009058 | 0.014248 |
| O18 | 0.634567 | 0.990942 | 0.985752 |

| O19 | 0.634567 | 0.009058 | 0.485752 |
| --- | --- | --- | --- |
| O20 | 0.365433 | 0.990942 | 0.514248 |
| O21 | 0.865433 | 0.509059 | 0.014247 |
| O22 | 0.134568 | 0.490941 | 0.985752 |
| O23 | 0.134567 | 0.509059 | 0.485752 |
| O24 | 0.865433 | 0.490941 | 0.514248 |

b) The ordered spin states of NaFeGe$_2$O$_6$.

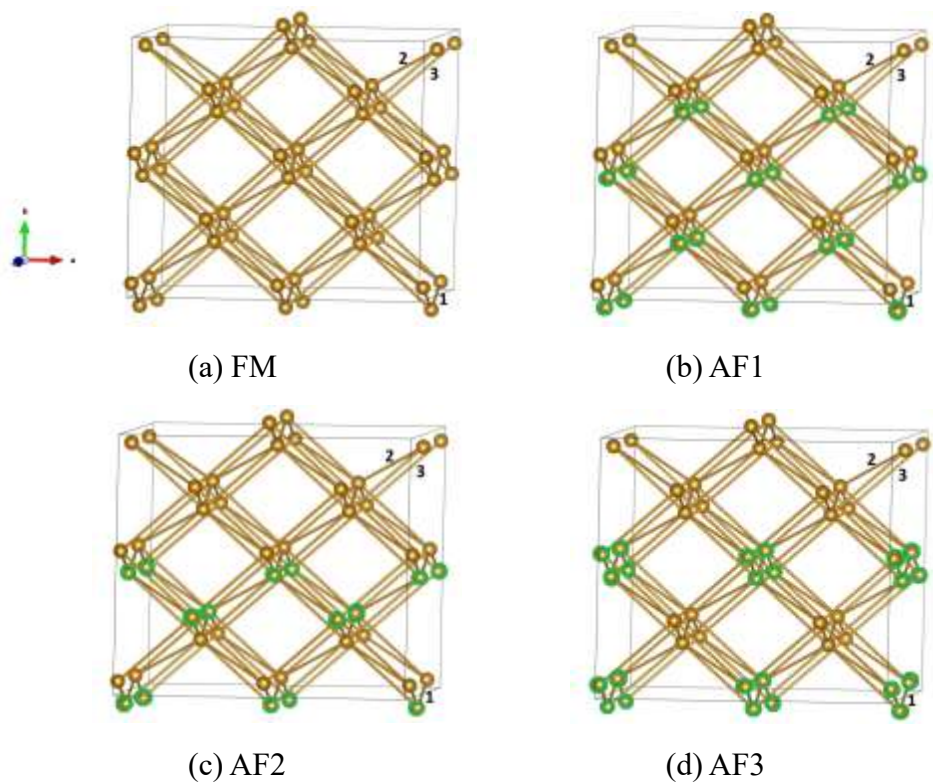

(a) FM         (b) AF1

(c) AF2        (d) AF3

**Fig. S7.** The ordered spin states used for NaFeGe$_2$O$_6$: (a) FM, (b) AF1, (c) AF2, and (d) AF3, where down-spin Fe$^{3+}$ sites are indicated by encircling them with green circles.

Table S6. The spin exchange energies for the ordered spin states of NaFeGe$_2$O$_6$.

$$E_{SE}(i) = \sum_{i=1}^{N} n_1(i) J_1 + n_2(i) J_2 + \cdots + n_N(i) J_N$$

| i | $n_1$ | $n_2$ | $n_3$ |
|---|---|---|---|
| FM | +4 | +4 | +8 |
| AF1 | -4 | -4 | +8 |
| AF2 | -4 | +4 | -8 |
| AF3 | +4 | -4 | -8 |

c) Table S7. The relative energies of the ordered spin states in terms of DFT+U calculations for the optimized structure of NaFeGe$_2$O$_6$.

| i | U = 3 eV | U = 4 eV |
|---|---|---|
| FM | 32.83 | 27.24 |
| AF1 | 0.29 | 0.25 |
| AF2 | 0 | 0 |
| AF3 | 16.80 | 14.02 |

d) Table S8. The values of the spin exchanges determined for NaFeGe$_2$O$_6$ by the energy mapping analysis.

|  | U = 3 eV | U = 4 eV |
|---|---|---|
| $J_1$ | 22.89 | 19.02 |
| $J_2$ | 7.30 | 6.02 |
| $J_3$ | 3.78 | 3.13 |

29## 3. LiFeGe₂O₆

a) Table S9. The optimized atom positions of LiFeGe$_2$O$_6$

|      | x        | y        | x        |
|------|----------|----------|----------|
| Fe1  | 0.748760 | 0.151354 | 0.288248 |
| Fe2  | 0.748760 | 0.348646 | 0.788248 |
| Fe3  | 0.251241 | 0.651354 | 0.211752 |
| Fe4  | 0.251241 | 0.848646 | 0.711752 |
| Li1  | 0.256693 | 0.492406 | 0.720205 |
| Li2  | 0.743307 | 0.507594 | 0.279795 |
| Li3  | 0.743307 | 0.992406 | 0.779795 |
| Li4  | 0.256693 | 0.007594 | 0.220205 |
| Ge1  | 0.047021 | 0.342075 | 0.282837 |
| Ge2  | 0.952979 | 0.657925 | 0.717163 |
| Ge3  | 0.952979 | 0.842075 | 0.217163 |
| Ge4  | 0.047021 | 0.157925 | 0.782837 |
| Ge5  | 0.553932 | 0.839625 | 0.231200 |
| Ge6  | 0.446068 | 0.160375 | 0.768800 |
| Ge7  | 0.446068 | 0.339625 | 0.268800 |
| Ge8  | 0.553932 | 0.660375 | 0.731200 |
| O1   | 0.857968 | 0.333999 | 0.180593 |
| O2   | 0.142032 | 0.666001 | 0.819407 |
| O3   | 0.142032 | 0.833999 | 0.319407 |
| O4   | 0.857968 | 0.166001 | 0.680593 |
| O5   | 0.112795 | 0.524543 | 0.284448 |
| O6   | 0.887205 | 0.475457 | 0.715552 |
| O7   | 0.887205 | 0.024543 | 0.215552 |
| O8  | 0.112795 | 0.975457 | 0.784448 |
| O9  | 0.121494 | 0.294688 | 0.622684 |
| O10 | 0.878506 | 0.705312 | 0.377316 |
| O11 | 0.878506 | 0.794688 | 0.877316 |
| O12 | 0.121494 | 0.205312 | 0.122684 |
| O13 | 0.363821 | 0.833879 | 0.107405 |
| O14 | 0.636179 | 0.166121 | 0.892595 |
| O15 | 0.636179 | 0.333879 | 0.392595 |
| O16 | 0.363821 | 0.666121 | 0.607405 |
| O17 | 0.631318 | 0.003772 | 0.393798 |
| O18 | 0.368682 | 0.996228 | 0.606202 |
| O19 | 0.368682 | 0.503772 | 0.106202 |
| O20 | 0.631318 | 0.496228 | 0.893798 |
| O21 | 0.614475 | 0.683925 | 0.454043 |
| O22 | 0.385525 | 0.316075 | 0.545957 |
| O23 | 0.385525 | 0.183925 | 0.045957 |
| O24 | 0.614475 | 0.816075 | 0.954043 |

b) The ordered spin states of LiFeGe$_2$O$_6$

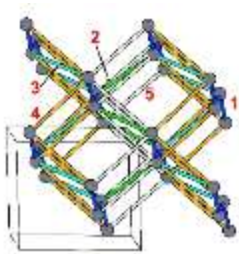 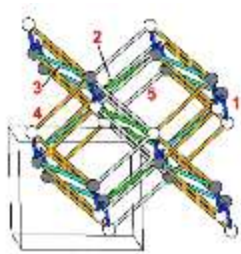 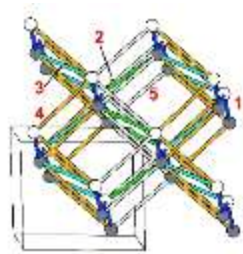

(a) FM     (b) AF1     (c) AF2

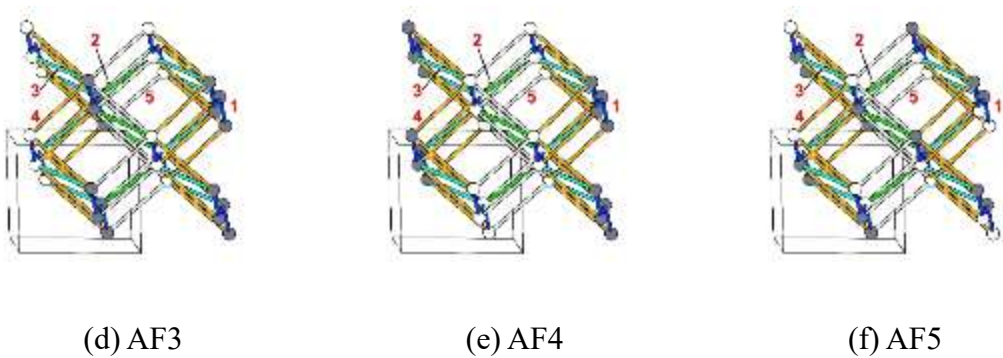

(d) AF3   (e) AF4   (f) AF5

**Fig. S8.** The ordered spin states used for LiFeGe$_2$O$_6$: (a) FM, (b) AF1, (c) AF2, (d) AF3, (e) AF4, and (f) AF5 states, where up-spin and down-spin Fe$^{3+}$ sites are indicated by shaded and unshaded circles, respectively.

Table S10. The spin exchange energies for the ordered spin states of LiFeGe$_2$O$_6$

$$E_{SE}(i) = \sum_{i=1}^{N} n_1(i) J_1 + n_2(i) J_2 + \cdots + n_N(i) J_N$$

| i | $n_1$ | $n_2$ | $n_3$ | $n_4$ | $n_5$ |
|---|---|---|---|---|---|
| FM | +32 | +16 | +16 | +32 | +32 |
| AF1 | -32 | +16 | +16 | -32 | -32 |
| AF2 | -32 | -16 | -16 | +32 | +32 |
| AF3 | +32 | -16 | -16 | -32 | -32 |
| AF4 | +32 | +16 | -16 | -32 | +32 |
| AF5 | -32 | +16 | -16 | +32 | -32 |

c) Table S11. The relative energies of the ordered spin states in terms of DFT+U calculations for the optimized structure of LiFeGe$_2$O$_6$.

| i | U = 3 eV | U = 4 eV |
|---|---|---|
| FM | 25.89 | 20.99 |
| AF1 | 2.13 | 1.82 |
| AF2 | 2.13 | 1.94 |
| AF3 | 12.22 | 9.80 |
| AF4 | 20.43 | 16.54 |
| AF5 | 0 | 0 |

d) Table S12. The values of the spin exchanges determined for LiFeGe$_2$O$_6$ by the energy mapping analysis.

| | U = 3 eV | U = 4 eV |
|---|---|---|
| J$_1$ | 15.71 | 12.54 |
| J$_2$ | 5.64 | 4.45 |
| J$_3$ | 7.05 | 5.81 |
| J$_4$ | 1.55 | 1.22 |
| J$_5$ | 4.80 | 4.03 |

4. SrMn$_2$V$_2$O$_8$

a) Table S13. The optimized atom positions of SrMn$_2$V$_2$O$_8$

| | x | y | z |
|---|---|---|---|
| Mn1 | 0.331929 | 0.333608 | 0.209992 |
| Mn2 | 0.333608 | 0.168071 | 0.459992 |
| Mn3 | 0.168071 | 0.166392 | 0.709992 |
| Mn4 | 0.166392 | 0.331929 | 0.959992 |
| Mn5 | 0.831929 | 0.166392 | 0.209992 |
| Mn6 | 0.833608 | 0.331929 | 0.459992 |
| Mn7 | 0.668071 | 0.333608 | 0.709992 |
| Mn8 | 0.666392 | 0.168071 | 0.959992 |
| Mn9 | 0.168071 | 0.833608 | 0.209992 |
| Mn10 | 0.166392 | 0.668071 | 0.459992 |
| Mn11 | 0.331929 | 0.666392 | 0.709992 |
| Mn12 | 0.333608 | 0.831929 | 0.959992 |
| Mn13 | 0.668071 | 0.666392 | 0.209992 |
| Mn14 | 0.666392 | 0.831929 | 0.459992 |
| Mn15 | 0.831929 | 0.833608 | 0.709992 |
| Mn16 | 0.833608 | 0.668071 | 0.959992 |
| Sr1 | 0.000000 | 0.000000 | 0.002656 |
| Sr2 | 0.500000 | 0.500000 | 0.502656 |
| Sr3 | 0.000000 | 0.500000 | 0.252656 |
| Sr4 | 0.500000 | 0.000000 | 0.752656 |
| Sr5 | 0.000000 | 0.000000 | 0.502656 |
| Sr6 | 0.500000 | 0.500000 | 0.002656 |
| Sr7 | 0.000000 | 0.500000 | 0.752656 |
| Sr8 | 0.500000 | 0.000000 | 0.252656 |
| V1 | 0.263785 | 0.079609 | 0.077522 |
| V2 | 0.236215 | 0.420391 | 0.577522 |
| V3 | 0.920391 | 0.763785 | 0.327522 |
| V4 | 0.579609 | 0.736215 | 0.827522 |
| V5 | 0.263785 | 0.920391 | 0.577522 |
| V6 | 0.236215 | 0.579609 | 0.077522 |
| V7 | 0.920391 | 0.236215 | 0.827522 |
| V8 | 0.579609 | 0.263785 | 0.327522 |
| V9 | 0.763785 | 0.579609 | 0.577522 |
| V10 | 0.736215 | 0.920391 | 0.077522 |
| V11 | 0.420391 | 0.263785 | 0.827522 |
| V12 | 0.079609 | 0.236215 | 0.327522 |
| V13 | 0.763785 | 0.420391 | 0.077522 |
| V14 | 0.736215 | 0.079609 | 0.577522 |
| V15 | 0.420391 | 0.736215 | 0.327522 |
| V16 | 0.079609 | 0.763785 | 0.827522 |
| O1 | 0.140655 | 0.498881 | 0.990198 |
| O2 | 0.359345 | 0.001119 | 0.490198 |
| O3 | 0.501119 | 0.640655 | 0.240198 |
| O4 | 0.998881 | 0.859345 | 0.740198 |
| O5 | 0.140655 | 0.501119 | 0.490198 |
| O6 | 0.359345 | 0.998881 | 0.990198 |
| O7 | 0.501119 | 0.359345 | 0.740198 |
| O8 | 0.998881 | 0.140655 | 0.240198 |
| O9 | 0.640655 | 0.998881 | 0.490198 |
| O10 | 0.859345 | 0.501119 | 0.990198 |
| O11 | 0.001119 | 0.140655 | 0.740198 |
| O12 | 0.498881 | 0.359345 | 0.240198 |
| O13 | 0.640655 | 0.001119 | 0.990198 |
| O14 | 0.859345 | 0.498881 | 0.490198 |
| O15 | 0.001119 | 0.859345 | 0.240198 |
| O16 | 0.498881 | 0.640655 | 0.740198 |
| O17 | 0.342861 | 0.667302 | 0.462947 |
| O18 | 0.157139 | 0.832698 | 0.962947 |
| O19 | 0.332698 | 0.842861 | 0.712947 |
| O20 | 0.167302 | 0.657139 | 0.212947 |
| O21 | 0.342861 | 0.332698 | 0.962947 |
| O22 | 0.157139 | 0.167302 | 0.462947 |
| O23 | 0.332698 | 0.157139 | 0.212947 |
| O24 | 0.167302 | 0.342861 | 0.712947 |
| O25 | 0.842861 | 0.167302 | 0.962947 |
| O26 | 0.657139 | 0.332698 | 0.462947 |
| O27 | 0.832698 | 0.342861 | 0.212947 |
| O28 | 0.667302 | 0.157139 | 0.712947 |
| O29 | 0.842861 | 0.832698 | 0.462947 |
| O30 | 0.657139 | 0.667302 | 0.962947 |
| O31 | 0.832698 | 0.657139 | 0.712947 |
| O32 | 0.667302 | 0.842861 | 0.212947 |
| O33 | 0.154479 | 0.685370 | 0.700527 |
| O34 | 0.345521 | 0.814630 | 0.200527 |
| O35 | 0.314630 | 0.654479 | 0.950527 |
| O36 | 0.185370 | 0.845521 | 0.450527 |
| O37 | 0.154479 | 0.314630 | 0.200527 |
| O38 | 0.345521 | 0.185370 | 0.700527 |
| O39 | 0.314630 | 0.345521 | 0.450527 |
| O40 | 0.185370 | 0.154479 | 0.950527 |
| O41 | 0.654479 | 0.185370 | 0.200527 |
| O42 | 0.845521 | 0.314630 | 0.700527 |
| O43 | 0.814630 | 0.154479 | 0.450527 |
| O44 | 0.685370 | 0.345521 | 0.950527 |
| O45 | 0.654479 | 0.814630 | 0.700527 |
| O46 | 0.845521 | 0.685370 | 0.200527 |
| O47 | 0.814630 | 0.845521 | 0.950527 |
| O48 | 0.685370 | 0.654479 | 0.450527 |
| O49 | 0.325153 | 0.499599 | 0.171844 |
| O50 | 0.174847 | 0.000401 | 0.671844 |
| O51 | 0.500401 | 0.825153 | 0.421844 |
| O52 | 0.999599 | 0.674847 | 0.921844 |
| O53 | 0.325153 | 0.500401 | 0.671844 |
| O54 | 0.174847 | 0.999599 | 0.171844 |
| O55 | 0.500401 | 0.174847 | 0.921844 |
| O56 | 0.999599 | 0.325153 | 0.421844 |
| O57 | 0.825153 | 0.999599 | 0.671844 |
| O58 | 0.674847 | 0.500401 | 0.171844 |
| O59 | 0.000401 | 0.325153 | 0.921844 |
| O60 | 0.499599 | 0.174847 | 0.421844 |
| O61 | 0.825153 | 0.000401 | 0.171844 |
| O62 | 0.674847 | 0.499599 | 0.671844 |
| O63 | 0.000401 | 0.674847 | 0.421844 |
| O64 | 0.499599 | 0.825153 | 0.921844 |

b) The ordered spin states of SrMn$_2$V$_2$O$_8$

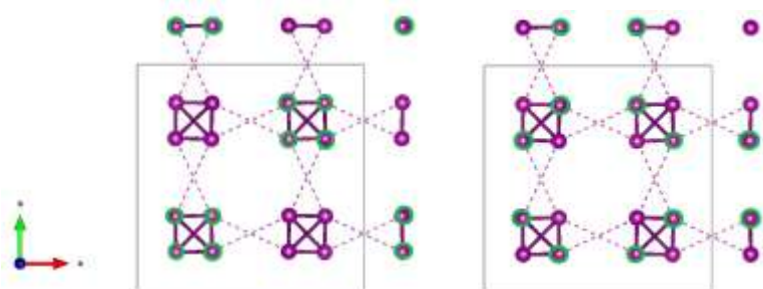

(a) AF1  (b) AF2

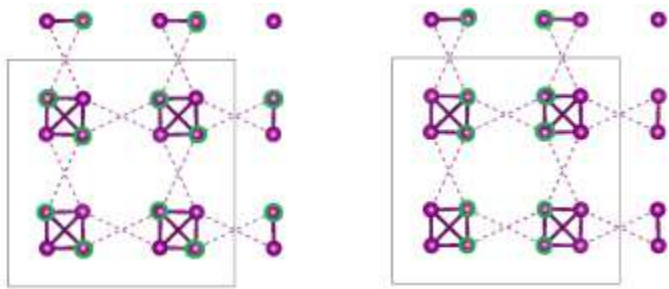

(c) AF3  (d) AF4

**Fig. S9.** The ordered spin states used for SrMn$_2$V$_2$O$_8$: (a) AF1, (b) AF2, (c) AF3, and (d) AF4 states, where up-spin Mn$^{2+}$ sites are indicated by encircling them with green circles.

Table S14. The spin exchange energies for the ordered spin states of SrMn$_2$V$_2$O$_8$

$$E_{SE}(i) = \sum_{i=1}^{N} n_1(i) J_1 + n_2(i) J_2 + \cdots + n_N(i) J_N$$

|     | $n_1$ | $n_2$ | $n_3$ |
|-----|-------|-------|-------|
| AF1 | +16   | +16   | -16   |
| AF2 | -16   | +16   | -16   |
| AF3 | -16   | +16   | +16   |
| AF4 | 0     | -16   | 0     |

c) Table S15. The relative energies of the ordered spin states in terms of DFT+U calculations for the optimized structure of SrMn$_2$V$_2$O$_8$.

| i   | U = 3 eV | U = 4 eV |
|-----|----------|----------|
| AF1 | 112.30   | 95.44    |
| AF2 | 0        | 0        |
| AF3 | 4.46     | 3.53     |
| AF4 | 54.02    | 45.98    |

d) Table S16. The values of the spin exchanges determined for SrMn$_2$V$_2$O$_8$ by the energy mapping analysis

|       | U = 3 eV | U = 4 eV |
|-------|----------|----------|
| $J_1$ | 52.11    | 44.28    |
| $J_2$ | 2.20     | 1.62     |

| J$_3$ | 2.07 | 1.64 |
|---|---|---|